\documentclass[12pt]{article}
\usepackage{latexsym}
\usepackage{amsmath,,calrsfs}
\usepackage{amsfonts}
\usepackage{amssymb}
\usepackage{amscd}
\usepackage{bbm}
\usepackage{fancybox}
\usepackage{cite}
\usepackage{amsmath,amsfonts,amsbsy}
\usepackage{pstricks,pst-node}
\usepackage[small,bf,hang]{caption2}
\usepackage{graphicx}
\usepackage{epsfig}
\usepackage{psfrag}
\usepackage{comment}

\usepackage{float}

\psset{unit=1.3cm,linewidth=.5pt,radius=.2}  

\usepackage{multirow}                     
\usepackage{float}                          
\usepackage{lscape}                         
\usepackage{bm}

\newcommand{\bb}{\bar\beta}

\newcommand{\beq}{\begin{equation}}
\newcommand{\eeq}{\end{equation}}
\newcommand{\bi}{\begin{itemize}}
\newcommand{\ei}{\end{itemize}}
\newcommand{\bt}{\begin{tabular}}
\newcommand{\et}{\end{tabular}}
\newcommand{\bc}{\begin{center}}
\newcommand{\ec}{\end{center}}

\newcommand{\be}{\begin{equation}}
\newcommand{\ee}{\end{equation}}
\newcommand{\bea}{\begin{eqnarray}}
\newcommand{\eea}{\end{eqnarray}}
\newcommand{\ba}{\begin{array}}
\newcommand{\ea}{\end{array}}

\def\bbox{{\,\lower0.9pt\vbox{\hrule \hbox{\vrule height 0.2 cm
\hskip 0.2 cm \vrule height 0.2 cm}\hrule}\,}}
\newcommand{\dsl}{\pa \kern-0.5em /}

\font\mybb=msbm10 at 12pt
\def\bb#1{\hbox{\mybb#1}}

\def\bI {\bb{I}}

\def\tr{{\rm tr}}

\begin{document}

\begin{titlepage}
\begin{center}

\hfill  DAMTP-1985-0263

\vskip 1.5cm

{\large \bf  The Jordan formulation of Quantum Mechanics: a review\footnote{From  ``Supersymmetry, Supergravity and Related Topics'', proceedings of the XVth GIFT International Seminar on Theoretical Physics,  Sant Feliu de Guixols, Girona, Spain, 4-9 June 1984; eds. F. del Aguila, J.A. de Azc\'arraga and L.E. Ib\'a\~nez, World Scientific 1985. }}

\vskip 1cm

{\bf Paul K.~Townsend} \\

\vskip 25pt

{\em  Department of Applied Mathematics and Theoretical Physics,\\ Centre for Mathematical Sciences, University of Cambridge,\\
Wilberforce Road, Cambridge, CB3 0WA, U.K.\vskip 5pt }

{email: {\tt P.K.Townsend@damtp.cam.ac.uk}} \\

\end{center}

\vskip 0.5cm

\begin{center} {\bf ABSTRACT}\\[3ex]
\end{center}

This is a transcription of a conference proceedings from 1985. It reviews the Jordan algebra formulation of quantum mechanics.  A possible novelty is the discussion of time 
evolution; the associator takes over the role of $i$ times  the commutator in the standard density matrix formulation,  and for the Jordan algebra of complex Hermitian 
matrices this implies a Hamiltonian of  the form $H= i[x,y] + \lambda \bI$ for traceless Hermitian $x,y$ and real number $\lambda$. Other possibilities for time evolution 
in the Jordan formulation are briefly considered.

\end{titlepage}

%


In the standard formulation of quantum mechanics an observable is represented by a (possibly infinite dimensional) Hermitian matrix. But if $a$ and $b$ are Hermitian the matrix product $ab$ is not necessarily Hermitian. Consider instead the Jordan product
\begin{equation}\label{Jprod}
a\circ b = \tfrac{1}{2}\left(ab+ba\right)\, , 
\end{equation}
which {\it is} Hermitian if $a$ and $b$ are. With respect to the Jordan product, Hermitian matrices form a closed algebra that is (obviously) commutative,  but  \emph{non-associative}. However, the Jordan product does have the property that
\begin{equation}\label{JI}
a\circ (a^2\circ b) = a^2 \circ (a\circ b)\, , 
\end{equation}
where $a^2= a \circ a$, which equals the matrix product of $a$ with itself. If $a$ and $b$ are genuine matrices (i.e. associative with respect to the matrix product) then this relation is an identity, known as the Jordan identity. 

Hermitian matrices also have the additional property that 
\begin{equation}
a\circ a + b\circ b =0 \quad \Rightarrow \quad a=b=0\, . 
\end{equation}
An algebra with this property is said to be \emph{formally real}. Given reality,  it can be shown that the Jordan identity is equivalent to \emph{power associativity}. As we have remarked, $a\circ a$ is unambiguously $a^2$. The product $a\circ a\circ a$ is also unambiguously  $a^3$  because of commutativity, but $a\circ a \circ a \circ a$ is potentially ambiguous as it could be $(a\circ a)\circ (a \circ a)$ or $a\circ(a\circ (a\circ a)$. Power associativity ensures that they agree and, more generally, that 
\begin{equation}
a^n\circ a^m = a^{m+n}
\end{equation}
for all positive integers $m$ and $n$. This property allows us to define unambiguously functions of an observable as power series expansions. We expect that if $\alpha$ is the value of $a$ in somestate then $f(\alpha)$ will be the value of $f(a)$ in that state. This is the physical motivation for power associativity and hence the Jordan identity. 

Of course, as long as we continue to regard the Jordan product as derived from the matrix product we have no need of motivations for identities such as (\ref{JI}); they are simply true. But we can \emph{define} the Jordan algebra directly in terms of the Jordan identity. More precisely, we define a Jordan algebra as a vector space with a commutative bilinear product satisfying the Jordan identity. To qualify as an \emph{algebra of observables} we require, in addition, that there be a unit element (because the unit matrix is Hermitian) and that the algebra be formally real. In this way we arrive at the axioms \cite{JA}
\begin{enumerate}

\item $a\circ b = b\circ a \qquad$ (commutativity)

\item $a\circ (a^2\circ b) = a^2\circ (a\circ b) \qquad$ (Jordan identity)

\item $a^2 + b^2 =0 \quad \Rightarrow\quad a=b=0\qquad$ (reality)

\item $\exists\  \bI$ such that $\bI\circ a= a \qquad$ (existence of identity element)

\end{enumerate}

Although Hermitian matrices certainly constitute an example of such an algebra, with the Jordan product as in (\ref{Jprod}), it is by no means obvious that they exhaust the possibilities. To investigate this point we need a classification of formally real unital Jordan algebras. This task is greatly simplified by the theorem that all non-simple Jordan algebras are direct sums of simple algebras. In the finite-dimensional case the classification was achieved in 1934 \cite{Jordan:1933vh}. There are five classes, which are 
\begin{enumerate}

\item  $J(Q)$

\item  $H_m^{\mathbb{R}}$

\item  $H_m^{\mathbb{C}}$

\item  $H_m^{\mathbb{H}}$

\item  $H_3^{\mathbb{O}}$

\end{enumerate}
Every element $a$ of a Jordan algebra can be written as
\begin{equation}
a = a^I e_I\, , 
\end{equation}
where $e_I=\{\mathbb{I}, e_i ; i=1, \dots,n\}$ is a basis for the algebra\footnote{In other words, as a vector space the algebra has dimension $n+1$.}, which can be chosen such that\footnote{The $e_i$ are the traceless matrices; the trace is discussed below.}
\begin{equation}
e_i\circ e_j = \delta_{ij} \mathbb{I} + T_{ijk} e_k\, . 
\end{equation}
In general, an arbitrary nonsingular matrix would replace $\delta_{ij}$ but reality implies that this may be transformed to $\delta_{ij}$ by a change of basis.  Similarly, one can take the structure constants $T_{ijk}$ to be totally symmetric in $ijk$ without loss of generality. 

The type-$1$ algebras are the simplest because $T_{ijk}$ vanishes. These are the generalizations of the algebra of Pauli matrices under anticommutation, and are therefore subalgebras of the Clifford algebras of positive definite quadratic forms ($Q$).  The type-$2,3,4$ algebras are realizations as Hermitian matrices over the real numbers, complex numbers and quaternions, respectively, with the Jordan product again being the anticommutator. By contrast, the single type-$5$ algebra is \emph{exceptional} because it has no such matrix realization. It can be realized by $3\times 3$ Hermitian ``matrices'' over the octonions but these are not matrices in the usual sense because they are not associative with respect to the matrix product. The exceptionality of $H_3^{\mathbb{O}}$ 
can be proved by means of the ``Glennie identity'',  which is eight-order in elements of the algebra \cite{McC}; if the elements are ordinary matrices then this is indeed an identity, 
but it is not satisfied by $H_3^{\mathbb{O}}$.  The classification of infinite-dimensional Jordan algebras has recently been accomplished. It appears that all infinite-dimensional simple Jordan algebras are extensions of types $1$ to $4$. In particular, there are no exceptional infinite-dimensional algebras \cite{McC}. 

The above classification already suggests one generalization of quantum mechanics; i.e. that for which the complex numbers are replaced by the real numbers, quaternions or octonions. It appears that real quantum mechanics is esentially equivalent to complex quantum mechanics, and quaternion complex mechanics suffers from a surplus of imaginary units and does not yield much that is new \cite{Finkelstein:1962tj}. The octonionic quantum mechanics based on the exceptional Jordan algebra \cite{Gunaydin:1978jq} is the most interesting case because exceptionality implies that no Hilbert space formulation is possible. This is therefore a genuine, and radical, generalization of quantum mechanics, although one without as yet any application.

But my purpose here is not to review the generalizations of quantum mechanics afforded by the Jordan algebra formulation but to explain how conventional quantum mechanics is contained in it. 
So far we have discussed the algebra of \emph{observables} but we have not discussed the representation of \emph{states}. In the usual formulation, states are identified with rays in a Hilbert space. But if $\{|k\rangle; k=1,2,\dots\}$ is a basis of normalized vectors in a Hilbert space then the Hermitian operators
\begin{equation}
E_k = |k\rangle \langle k|\, , \quad  k=1,2,\dots
\end{equation}
contain the same information. These operators satisfy
\begin{equation}\label{normalize}
\tr\,  E =1\, , 
\end{equation}
and
\begin{equation}
E^2=E\, . 
\end{equation}

To carry this over to the Jordan formulation, we have first to introduce the \emph{trace form} of a Jordan algebra, which is defined to satisfy
\begin{equation}
\tr \, \bI = \nu\, , \qquad \tr \, e_i=0\, , 
\end{equation}
with $\nu$ an integer to be specified shortly. This is equivalent to the introduction of the positive definite inner product
\begin{equation}
\left(e_I,e_J\right) := \frac{1}{\nu} \, \tr\left(e_I\circ e_J\right) = \delta_{IJ}\, . 
\end{equation}
Now, an element $P$ of an algebra satisfying $P^2=P$ is called an \emph{idempotent}. For idempotents $P_1,P_2$, orthogonality with respect to the above inner product implies the seemingly stronger relation
\begin{equation}
P_1 \circ P_2 =0\, , 
\end{equation}
which we may take as the definition of orthogonality for idempotents. An idempotent $E$ that cannot be expressed as the sum of two orthogonal idempotents is said to be \emph{primitive}. The maximal number of mutually orthogonal primitive idempotents is the \emph{degree} of the Jordan algebra, and any such  set provides a decomposition of unity:
\begin{equation}
\sum_{\alpha=1}^{\rm deg} E_\alpha =1\, . 
\end{equation}
If the trace form is normalized as in (\ref{normalize}) then the \emph{degree} is the integer $\nu$. Thus, a Jordan algebra of  degree $\nu$ is the algebra of observables for  a quantum system with 
$\nu$ linearly independent states.  Accoding to a theorem of Jordan, Von Neumann and Wigner, any observable $a$ can be expressed as 
\begin{equation}
a = \sum_{\alpha=1}^\nu a_\alpha E_\alpha(a)\, , 
\end{equation}
i.e. as an expansion on a complete set of primitive idempotents, the set depending on the element chosen. The numbers $a_\alpha$ are the ``eigenvalues'' of the observable $a$ in the state represented by $E_\alpha(a)$. 

The primitive idempotents represent \emph{pure} states. A mixed state is represented by
\begin{equation}
\rho = \sum_{\alpha=1}^\nu p_\alpha E_\alpha \, , \qquad \sum_{\alpha=1}^\nu p_\alpha =1 \, , \quad p_\alpha\ge0 \,  \quad (\alpha=1,\dots,\nu)
\end{equation}
i.e. by a positive definite element $\rho$ satisfying $\tr\,  \rho =1$ but $\rho^2 \ne\rho$. The expectation value of any observable $a$ in the  state 
represented by $\rho$ is 
\begin{equation}
\langle a \rangle = \tr (a\circ \rho) \, . 
\end{equation}
This is essentially the density matrix formalism of the standard formulation of quantum mechanics. 

A distinction between the Jordan formulation and the standard density matrix formulation of quantum mechanics arises when one considers time evolution. In the density matrix formalism one postulates the unitary time evolution equation 
\begin{equation}\label{unitary}
\dot \rho = -i \left[H,\rho\right]\, , 
\end{equation}
where $H$ is the (usually positive definite) Hermitian Hamiltonian. This equation has the property that it takes pure states into pure states. If one requires this property then 
the time evolution equation must take the following form in the Jordan formulation:
\begin{equation}\label{evolve}
\dot\rho = D(\rho)\, , 
\end{equation}
where $D$ is a linear operator with the property that 
\begin{equation}
D\left(e_I\circ e_J\right) = D(e_I) \circ e_J + e_I \circ D(e_J)\, , 
\end{equation}
i.e. D is required to be a \emph{derivation} of the Jordan algebra. The derivations form a Lie algebra under commutation and they generate the automorphism group of the algebra. It is a remarkable fact that the derivations of a Jordan algebra can all be expressed as 
\begin{equation}
D_{x,y} = \left[ L_x,L_y\right]\, , 
\end{equation}
where $L_x$ is the operation of multiplication by a traceless element $x$. The Jacobi identity is satisfied as a consequence of the Jordan identity for $x$ and $y$. This representation of the 
derivations allows us to rewrite  (\ref{evolve}) as
\begin{equation}\label{evolve2}
\dot\rho = D_{x,y}(\rho) = x\circ (\rho \circ y) - (x\circ \rho)\circ y\, . 
\end{equation}
Defining the \emph{associator} of three elements $a,b,c$ as 
\begin{equation}
\left[a,b,c\right] := a\circ(b\circ c) - (a\circ b)\circ c\, , 
\end{equation}
we see that (\ref{evolve2}) is 
\begin{equation}\label{evolve3}
\dot\rho = \left[x,\rho,y\right] \, . 
\end{equation}

The associator plays a role in the Jordan formulation of quantum mechanics analogous to the role of the commutator in the density matrix formulation. To make contact with the latter we work out the right hand side of (\ref{evolve3})  for complex Hermitian matrices $x,y$ and $\rho$, with Jordan product (\ref{Jprod}). One finds that 
\begin{equation}
\dot\rho =  [[x,y], \rho]\, ,
\end{equation}
so that we have equivalence with (\ref{unitary}) if 
\begin{equation}
H= i[x,y] + \lambda \, \bI\, , 
\end{equation}
for any real constant $\lambda$. Thus, in a sense, the Jordan formulation is the ``square root'' of the standard formulation; one wonders whether there is a connection to 
supersymmetry here. The interpretation of the traceless elements $x$ and $y$ in the case of real, quaternionic or octonionic Jordan algebras is problematic. 

Density matrices belong to the cone of positive definite Hermitian matrices. This has a generalization to any Jordan algebra $J$ and the cone $C(J)$ is called its domain of positivity. The group of linear transformations of coordinates of a cone is its automorphism group, and if this group acts transitively the cone is said to be homogeneous. A domain of positivity $C(J)$ is a 
homogeneous cone  and its automorphism group  is generated by the derivations of $J$ together with the operation $L$ of multiplication by an element of $J$. The derivations form a subalgebra that generate the stability group of the cone, and the linear operators $L$ generate translations in the cone. 

If one considers the possibility of a more general time evolution for which pure states do not necessarily evolve into pure states \cite{Hawking:1976ra} then one might think to generalize (\ref{evolve}) to include a term $L(\rho)$ in the expression for $\dot\rho$. But this leads to a violation  the unit trace condition on $\rho$. One can construct an operation that preserves the trace by rescaling $\rho$ but this is not a linear operation because the rescaling factor is $\rho$-dependent. Thus,  one cannot use the automorphism group of the cone  to evolve density matrices into density matrices via a \emph{linear} evolution equation (except, of course, for the stability subgroup that evolves pure states into pure states). Conversely, those \emph{linear} transformations  that take pure states into mixed states cannot form a group. This is in accord with the second law of thermodynamics, which allows such transformations to form a semi-group but not a group, but the precise formulation of such generalized evolution equations in Jordan algebraic language has yet to be worked out.

\section*{Acknowledgements} I am grateful to  Gary Gibbons, German Sierra and Murat G\"unaydin for helpful conversations and correspondence.


\begin{thebibliography}{1}


\bibitem{JA}
P. Jordan, ``Uber Verallgemeinerungsm\"oglichkeiten des Formalismus der Quanten-mechanik'', {\it Nachr. Ges. Wiss. G\"ottingen},  (1933), 209

  
\bibitem{Jordan:1933vh}
  P.~Jordan, J.~von Neumann and E.~P.~Wigner,
  ``On an Algebraic generalization of the quantum mechanical formalism'',
  Annals Math.\  {\bf 35} (1934) 29.
  

\bibitem{McC}
K. McCrimmon, ``The Russian revolution in Jordan algebras'', Algebras, Groups \& Geometries ${\bf 1}$, (1984) 1. 

\bibitem{Finkelstein:1962tj}
  D.~Finkelstein, J.~M.~Jauch, S.~Schminovich and D.~Speiser,
  ``Principle of general Q covariance'',
  J.\ Math.\ Phys.\  {\bf 4} (1963) 788.
  
\bibitem{Gunaydin:1978jq}
  M.~G\"unaydin, C.~Piron and H.~Ruegg,
  ``Moufang Plane and Octonionic Quantum Mechanics'',
  Commun.\ Math.\ Phys.\  {\bf 61} (1978) 69.
  
\bibitem{Hawking:1976ra}
  S.~W.~Hawking,
  ``Breakdown of Predictability in Gravitational Collapse'',
  Phys.\ Rev.\ D {\bf 14} (1976) 2460.



\end{thebibliography}

\providecommand{\href}[2]{#2}\begingroup\raggedright\endgroup

\end{document}